\documentstyle[osa,aplop,manuscript]{revtex}
\def\i{{\rm i}}
\def\d{{\rm d}}
\def\e{{\rm e}}
\def\vector#1{{\bf #1}}
\def\vq{{\vector q}}
\def\vr{{\vector r}}
\def\vvF{{\vector v}_{\rm F}}
\def\vphat{{\hat {\vector p}}}
\def\dps{\displaystyle}
\def\vF{{v_{\rm F}}}
\def\Tc{{T_{\rm c}}}
\def\qbar{{\bar q}}
\def\hightc{{high-$T_{\rm c}$ }} 
\def\Tczero{{T_{\rm c}^{(0)}}}
\def\lesssim{\stackrel{{\textstyle<}}{\raisebox{-.75ex}{$\sim$}}}

\title{
Structure of the Fulde-Ferrell-Larkin-Ovchinnikov state 
in two-dimensional superconductors }

\author{Hiroshi Shimahara}

\address{
Faculty of Science, Hiroshima University, 
Higashi-Hiroshima 739, Japan
}

\date{(Received ~~~~~~~~~ 1997)}

\begin{document}

\maketitle

\begin{abstract}
Nonuniform superconducting state due to strong spin magnetism 
is studied in two-dimensional (2D) type-II superconductors 
near the second order phase transition line between 
the normal and the superconducting states. 
The optimum spatial structure of the orderparameter is examined 
in systems with cylindrical symmetric Fermi surfaces. 
It is found that states with 2D structures have lower 
free energies than the traditional one-dimensional solutions, 
at low temperatures and high magnetic fields. 
For $s$-wave pairing, triangular, square, hexagonal states are favored 
depending on the temperature, while square states are favored at low 
temperatures for $d$-wave pairing. 
In these states, orderparameters have 2D structures 
such as square and triangular lattices. 
\end{abstract}

\pacs{
PACS numbers: 74.60.-w,74.70.Kn,74.72.-h,74.76.-w} 

\newpage

Nonuniform superconducting state due to a strong spin magnetic 
effect, which is called Fulde-Ferrell-Larkin-Ovchinnikov (FFLO) state, 
has often been studied, since Fulde \& Ferrell and 
Larkin \& Ovchinnikov showed that the upper critical field of 
a nonuniform superconducting state exceeds the Pauli paramagnetic limit 
(Chandrasekar-Clogston limit) under some ideal 
conditions.~\cite{ful64,lar64,gru66,bul73,aoi74,buz83,suz83,mac84,leb86,buz87,DMM,bur94,Q2D,dup95,dw,BEDT,buz96,tiltH,glo93,sch94,yin93,tak69,psm,NUSDl,tac96,NUSDfp,buz96b,mat97}
Recently, it is sometimes suggested that organic, \hightc copper oxide, 
and heavy fermion superconductors can be candidates of the FFLO 
superconductors.~\cite{buz83,suz83,mac84,leb86,buz87,DMM,bur94,Q2D,dup95,dw,BEDT,buz96,tiltH,glo93,sch94,yin93}
This is because they have large upper critical fields so that 
strong spin magnetism is attainable, 
and also because they can be clean type-II superconductors.

In particular, the organic and the copper oxide superconductors have 
large critical fields, especially when orbital pair breaking effect is 
weaken by applying magnetic field in any direction parallel to the most 
conducting plane.~\cite{leb86}
In addition to this, some features of Fermi surface structure 
arising from the low dimensionality enhances the critical field of 
the FFLO state.~\cite{bul73,aoi74,buz83,suz83,mac84,leb86,buz87,DMM,bur94,Q2D,dup95,dw,BEDT,buz96,tiltH}
Therefore, some compounds in the families of the organic and \hightc 
superconductors might be good FFLO superconductors.

However, spatial oscillation of the orderparameter, 
which characterizes the FFLO state, has not been observed yet, 
while very large upper critical fields were observed in some compounds 
such as ${\rm (BEDT}$-${\rm TTF)_2X}$ and ${\rm (TMTSF)_2X}$. 
Those critical fields might exceed the Pauli paramagnetic 
limit.~\cite{PF6,mur88,betaET,5fold,ETI3} 
In particular, increase of the critical field at low temperatures 
with positive $\d^2 H_c/\d T^2$ was observed 
in ${\rm (TMTSF)_2PF_6}$.~\cite{PF6}
Such behavior of the critical field is very similar to the behavior 
that is theoretically obtained in low dimensional FFLO 
superconductors.~\cite{bul73,aoi74,buz83,suz83,mac84,leb86,buz87,DMM,bur94,Q2D,dup95,dw,BEDT,buz96,tiltH}
This might suggest the possibility of the FFLO state in this material 
or similar organic compounds, although this is not an evidence of 
the FFLO state. 
The spatial structure of the orderparameter must be observed 
in order to prove existence or nonexistence of the FFLO state.

The spatial structure of the orderparameter in the FFLO state is 
different from that in the traditional Abrikosov state, because 
their origins are quite different. 
The nonuniformity of the orderparameter in the FFLO state is due to 
the displacement of the Fermi surfaces of up and down spin electrons 
by the Zeeman energy, not due to the orbital motion around flux lines. 
Gruenberg and Gunther examined the condition for the coexistence 
of these states in a three dimensional system.~\cite{gru66} 
In the two dimensions, the Abrikosov state occurs unless the magnetic 
field is parallel to the conducting layer. 
However, when the parallel direction is approached, 
its Landau level index increases,~\cite{bul73,buz96,tiltH} 
and then the spatial structure of the orderparameter approaches to that 
of the FFLO state.~\cite{tiltH} 
We assume a parallel magnetic field and concentrate 
on the FFLO state in this paper.

In the studies so far, the spatial oscillation of the orderparameter 
in the FFLO state is believed to be in a single direction. 
Larkin {\it et al.} studied an $s$-wave superconductor 
with spherical symmetric Fermi surfaces, and found that the form 
$\Delta({\vr}) \propto \cos(\vq \cdot \vr)$ is the most stable 
solution among the periodic solutions of the gap equation 
expanded near the second order phase transition field.~\cite{lar64} 
However, other cases such as systems with anisotropic 
Fermi surfaces and those with anisotropic pairing interactions 
have not been studied yet. 
In particular, two-dimensional (2D) model is significant when one 
considers the organic and copper oxide superconductors. 
Orderparameters with spatial structures such as the square lattice and 
triangular lattice becomes small in larger area than the orderparameter 
oscillating in a single direction. 
Thus, the states with such 2D lattice structures 
would gain more spin-polarization energy 
than the state with the one-dimensional (1D) structures, 
and have chance to occur at high magnetic fields. 
Hence, such states may occur in the 2D FFLO 
superconductors (including quasi-1D systems),
since the critical field of the FFLO state remarkably increases at low 
temperatures.~\cite{bul73,aoi74,buz83,suz83,mac84,leb86,buz87,DMM,bur94,Q2D,dup95,dw,BEDT,buz96,tiltH}

In this paper, we study 2D FFLO superconductors 
with cylindrical symmetric Fermi surfaces. 
We examine $s$-wave pairing and $d$-wave pairing, the latter which 
is a serious candidate of the organic and \hightc superconductors. 
Our theory is a straightforward extension of the work by Larkin 
{\it et al.} to the 2D systems, finite 
temperatures, and anisotropic superconductivity. 
We consider the magnetic field parallel to the conducting layers, 
and neglect the orbital pair breaking effect, although weak interlayer 
interactions are implicitly assumed so that the mean field treatment 
like BCS theory is justified at low temperatures.

At first, we examine the $s$-wave pairing. 
As many authors studied, 
a gap function of the generalized form 
\def\eqgapFF
{(1)}
$$
     \Delta(\vr) = \Delta_{\vq} \e^{\i \vq\cdot\vr} 
     \eqno\eqgapFF
     $$
has the highest second order transition field. 
The optimum value of $|\vq|$ is finite for 
$T < T^{*} \approx 0.56 \Tczero$, where $\Tczero$ is the zero field 
transition temperature. 
For example, the optimum value of $|\vq|$ is equal 
to $2h/\vF$ at $T=0$, 
where $h = \mu_0 H$ with the electron magnetic moment $\mu_0$. 
In the present case, because of the cylindrical symmetry of the system, 
any linear combinations 
\def\eqgapFFLO
{(2)}
$$
     \Delta(\vr) = \sum_{m} \Delta_m \e^{\i \vq_m\cdot\vr} , 
     \eqno\eqgapFFLO
     $$
have the same critical field, where $\vq_m$'s are any vectors of 
the optimum magnitude $|\vq_m|=q$ parallel to the conducting layer. 
However, such degeneracy is removed by the nonlinear terms in the gap 
equation, apart from the rotation as a whole: a particular kind of 
linear combinations have a lower free energy than the others, 
just below the critical field. 
Here, we have taken a finite number of $\vq_m$'s in the above, 
because we are considering periodic solutions. 
Near the transition point, the gap equation is expanded as 
\def\eqgapeq
{(3)}
$$
    \begin{array}{rcl} 
    \lefteqn{ 
    \dps{ \log(\frac{T_{\rm c}}{T}) \Delta_l^{*} } 
    = \dps{ \sum_m [   (2-\delta_{lm}) J(\theta_{lm}) 
                     \Delta_m^{*} \Delta_m \Delta_l^{*}       } } \\
    && \dps{     + (1 - \delta_{lm} - \delta_{l,-m}) 
                 {\tilde J}(\theta_{lm}) 
                     \Delta_m^{*} \Delta_{-l} \Delta_{-m}^{*}  ] , }
    \end{array}
    \eqno\eqgapeq
    $$
where $\theta_{lm}$ is the angle between $\vq_l$ and $\vq_m$, and 
\def\eqJ
{(4)}
$$
    \begin{array}{rcl}
    J(\theta_{lm}) & = & \dps{
             T \sum_n \int^{\infty}_{-\infty} \!\!\!\!\!\! \d \xi
             \int^{2\pi}_{0} \frac{\d \varphi}{2\pi}  
                (\i \omega_n + \xi + h)^{-2}   }\\
     &&  \times (\i \omega_n - \xi - \vvF\cdot\vq_m + h)^{-1}   \\
     &&  \times (\i \omega_n - \xi - \vvF\cdot\vq_l + h)^{-1}  
    \end{array}
    \eqno\eqJ
    $$
and 
\def\eqJtilde
{(5)}
$$
    \begin{array}{rcl}
    {\tilde J}(\theta_{lm}) 
    & = & \dps{  T \sum_n \int^{\infty}_{-\infty} \!\!\!\!\!\! \d \xi
               \int^{2\pi}_{0} \frac{\d \varphi}{2\pi}
              (\i \omega_n + \xi + h)^{-1}                       }\\
    && \times (\i \omega_n - \xi + h - \vvF\cdot\vq_l)^{-1}       \\
    && \times (\i \omega_n + \xi + h + \vvF\cdot\vq_l 
                                     + \vvF\cdot\vq_m)^{-1}       \\
    && \times (\i \omega_n - \xi + h - \vvF\cdot\vq_m)^{-1} .
    \end{array}
    \eqno\eqJtilde
    $$

We examine periodic solutions 
\renewcommand{\arraystretch}{1.2}
\def\eqDelta
{(6)}
$$
    \begin{array}{lrcl}
    \mbox{(a)} & \Delta(\vr) & = & \Delta_{\rm FF} \exp(\i \vq \cdot \vr)  \\
    \mbox{(b)} & \Delta(\vr) & = & 2 \Delta_{\rm FFLO} \cos(\vq \cdot \vr) \\
    \mbox{(c)} & \Delta(\vr) & = & 
      2 \Delta_{\rm sq} [\cos(q x) + \cos(q y)] \\
    \mbox{(d)} & \Delta(\vr) & = & 
        \Delta_{\rm tri} 
                 [\exp(\i \vq_1 \! \cdot \! \vr) 
                + \exp(\i \vq_2 \! \cdot \! \vr) \\
       &&&~~~~~ + \exp(\i \vq_3 \! \cdot \! \vr)] \\
    \mbox{(e)} & \Delta(\vr) & = & 
      2 \Delta_{\rm hexa}
                 [\cos(\vq_1 \! \cdot \! \vr) 
                + \cos(\vq_2 \! \cdot \! \vr) \\
       &&&~~~~~ + \cos(\vq_3 \! \cdot \! \vr)] 
    \end{array}
    $$
where $\vq_1 = q (1, 0)$, $\vq_2 = q (-1/2, \sqrt{3}/2)$, and 
$\vq_3 = q (-1/2, -\sqrt{3}/2)$. 
We refer to the states expressed by (a), (b), (c), (d), and (e) 
as FF state, traditional FFLO state, square state, triangular state, 
and hexagonal state, respectively.

We define the factor $a_{\alpha}$ 
\def\eqAdef
{(7)}
$$
    |\Delta_\alpha|^2 = a_{\alpha} \frac{\Tc - T}{\Tc} , 
    \eqno\eqAdef
    $$
where $\alpha$ indicates the type of state. 
For example, $\alpha$ is replaced with 
${\rm FF}, {\rm FFLO}, {\rm sq}, \cdots$, 
for the states ${\rm (a)}, {\rm (b)}, {\rm (c)}, \cdots$, respectively. 
The free energy per unit volume is calculated by the formula 
\def\eqfreeenergy
{(8)}
$$
    \Omega - \Omega_0 = - \frac{1}{2} N(0) \frac{\Tc - T}{\Tc} 
      \frac{1}{V} \int \d^3 \vr |\Delta(\vr)|^2 . 
    \eqno\eqfreeenergy
    $$
We define the factor $b_{\alpha}$ by 
\def\eqBdef
{(9)}
$$
    \Omega - \Omega_0 = - \frac{1}{2} N(0) b_{\alpha} 
      [\frac{\Tc - T}{\Tc}]^2 . 
    \eqno\eqBdef
    $$
From the gap equation, we have 
\def\eqBresults
{(10)}
$$
    \begin{array}{rcl}
    b_{\rm FF}   & = & \dps{ 1/J(0) }\\
    b_{\rm FFLO} & = & \dps{ 2/[J(0) + 2 J(\pi)] }\\
    b_{\rm sq} & = & \dps{ 4/[J(0) + 2 J(\pi) + 4 J(\pi/2) }\\
                   && \dps{            + 2 {\tilde J}(\pi/2) ] }\\
    b_{\rm tri}  & = & \dps{ 3/[J(0) + 4 J(2\pi/3) ] }\\
    b_{\rm hexa} & = & \dps{ 6/[J(0) + 2 J(\pi) + 4J(\pi/3) }\\
                 && \dps{       + 4 J(2\pi/3) 
                                + 2 {\tilde J}(\pi/3) 
                                + 2 {\tilde J}(2\pi/3) ] } . 
    \end{array}
    \eqno\eqBresults
    $$
In the limit of $T \rightarrow 0$, we have 
$J(0) = -1/[4h^2(1-\qbar^2)^{3/2}]$ for $\qbar < 1$, and $J(0) = 0$ 
for $\qbar > 1$, where $\qbar = \vF q/(2h)$. 
Since the optimum value of $\qbar$ is equal to 1 at $T=0$, 
the expansion factor $J(0)$ diverges. 
Thus, the gap equation can not be expanded in the power of $\Delta$ 
in two-dimensions at $T=0$. 
It is easily verified that there is a term proportional to 
$\Delta \log\Delta$ in the asymptotic form.

For finite temperatures, we estimate $b_{\alpha}$ numerically. 
The results are shown in Fig.\ref{fig:swave}. 
It is found that the solution of the form (b), 
{\it i.e.}, the traditional FFLO state, 
is optimum for $0.24 \Tczero \lesssim T < T^{*} \approx 0.56 \Tczero$. 
Below $T \approx 0.24 \Tczero$, however, other solutions have 
lower free energies. 
The triangular, square, hexagonal states are optimum 
for $0.16 \Tczero \lesssim T \lesssim 0.24 \Tczero$, 
$0.05 \Tczero \lesssim T \lesssim 0.16 \Tczero$, and 
$T \lesssim 0.05 \Tczero$, respectively. 
This result is plausible because when the temperature decreases and 
the magnetic field increases, 
node of the orderparameter in the real space becomes more favorable 
owing to gain in the spin-polarization energy. 
Our result is consistent with the result by Buzdin et al., in which 
the structure of the orderparameter is studied near the tricritical 
point.~\cite{BuzKach}

It is easily verified that the orderparameters of the triangular and 
hexagonal states have spatial structures of the triangular lattice, 
while that of the square state have a structure of the square lattice. 
The periodicity of the triangular lattice of $|\Delta(\vr)|$ in the 
hexagonal state is twice that in the triangular state.

It is easy to extend the theory to anisotropic superconductivities. 
The gap function has the form 
\def\eqgapFFdwave
{(11)}
$$
     \Delta(\vphat,\vr) = \sum_{m} \Delta_{m} 
                          \gamma(\vphat) 
                          \e^{\i \vq_m\cdot\vr} 
     \eqno\eqgapFFdwave
     $$
where $\gamma(\vphat) = \sqrt{2}({\hat p}_x^2 - {\hat p}_y^2)$ for 
$d$-wave pairing. 
As Maki {\it et al.} studied, the angle between the optimum vector $\vq$ 
and the $x$-axis is equal to $n\pi/2$ for low temperatures 
$T/\Tczero \lesssim 0.06$, 
while $\pi/4+n\pi/2$ for high temperatures 
$0.06 \lesssim T/\Tczero \lesssim 0.56$, 
where $n$ is integer.~\cite{dw} 
Since there are four equivalent optimum directions of $\vq$, 
we have three candidates to examine, {\it i.e.}, 
(a) the FF state, (b) the traditional FFLO state, 
and (c) the square state.

In this case, a factor $[\gamma(\vphat)]^4$ is introduced into the 
integrands of eq.{\eqJ} and eq.{\eqJtilde}. 
Free energies of the above three states are compared in 
Fig.\ref{fig:dwavelt} and Fig.\ref{fig:dwaveht}. 
Figure \ref{fig:dwavelt} shows the results for the low temperature 
phase, which is realized for $T/\Tczero \lesssim 0.06$, 
while Fig.\ref{fig:dwaveht} shows the results for the high temperature 
phase, which is realized for $0.06 \lesssim T/\Tczero \lesssim 0.56$. 
In Fig.\ref{fig:dwavelt}, we find that the square state is optimum 
for $T/\Tczero \lesssim 0.06$. 
In Fig.\ref{fig:dwaveht}, it is found that the square state is optimum 
for $0.085 \lesssim T/\Tczero \lesssim 0.12$, 
while the traditional FFLO state is optimum for 
$0.12 \lesssim T/\Tczero \lesssim 0.56$. 
At $T/\Tczero \approx 0.085$, the factors $a_{\rm sq}$ 
and $b_{\rm sq}$ turn negative. 
Thus, the gap equation expanded up to the third order does not have 
any solution of this type for $0.06 \lesssim T/\Tczero \lesssim 0.085$ 
below and near the transition temperatures. 
This suggests that a first order transition to the square FFLO state 
occurs at a field higher than the second order transition field, 
in this temperature region, as discussed below.

If the gap equation is simply derived from a differentiation of 
a free energy functional expanded in the power of $|\Delta_{\rm sq}|$, 
the factor of $[\Delta_{\rm sq}]^4$ in the free energy is negative 
near the second order transition point in this temperature region, 
since it is proportional to the inverse of $a_{\rm sq}$ 
and that of $b_{\rm sq}$. 
Thus, the free energy as a function of $|\Delta_{\rm sq}|$ 
has a minimum at nonzero $|\Delta_{\rm sq}|$ 
at the second order transition point, 
because there must be higher order terms such as a one proportional 
to $|\Delta_{\rm sq}|^6$, which guarantee that the free energy 
increases as a function of $|\Delta_{\rm sq}|$ where 
$|\Delta_{\rm sq}|$ is large. 
This finite value of $|\Delta_{\rm sq}|$ is the optimum 
solution of the gap equation at this point, 
and this solution must terminate at a higher field. 
Therefore, it is plausible that the actual critical field is higher 
than the second order transition field, and the phase transition 
is of the first order. 

A similar conjecture as the above holds for three-dimensional 
systems. As Larkin {\it et al.} suggested a possibility of 
the first order transition to the cubic state,~\cite{lar64} 
the same possibilities for the triangular state and the square state 
can not be excluded, 
because the factors $a_{\alpha}$ for these states are negative in the 
spherical symmetric case.

Lastly, we breifly discuss orbital pair breaking effect 
in three-dimensional FFLO superconductors (including quasi-2D systems). 
For the 1D solution of the traditional FFLO state, 
since the vector $\vq$ can be oriented in the direction of the magnetic 
field, the FFLO state can coexist with the vortex state if orbital 
effect is sufficiently weak.~\cite{gru66} 
On the other hand, for the present 2D solutions, 
the competition with the vortex states may be a crutial problem, 
since the vortex functions take two out of three digrees of freedom 
of the coordinate space. 
Lebed argued that if the magnetic field is sufficiently strong, 
the quasi-2D system can be treated as being essentially 
purely 2D.~\cite{leb86}
In such a case, the competition mentioned above does not occur 
if the magnetic field is accurately parallel to the conducting 
plane.~\cite{tiltH}

In conclusion, we have examined FFLO superconductors with the 
cylindrical symmetric Fermi surfaces near the second order transition 
line, and have found that spatial structures of the 2D 
lattices are favored more than the 1D structure of the 
traditional FFLO state, for low temperatures and high magnetic fields, 
both for the $s$-wave pairing and the $d$-wave pairing. 
It is found that 
the traditional FFLO state, the triangular state, the square state, 
and the hexagonal state appear for the $s$-wave pairing 
as the temperature decreases, 
while the square states occur at low temperatures 
for the $d$-wave pairing. 
If such structures of orderparameters are observed experimentally, 
for example, directly by the scanning tunneling microscope (STM) 
technique, it can be regarded as a strong evidence of existence 
of a nonuniform superconducting state.

\renewcommand{\baselinestretch}{1.0}


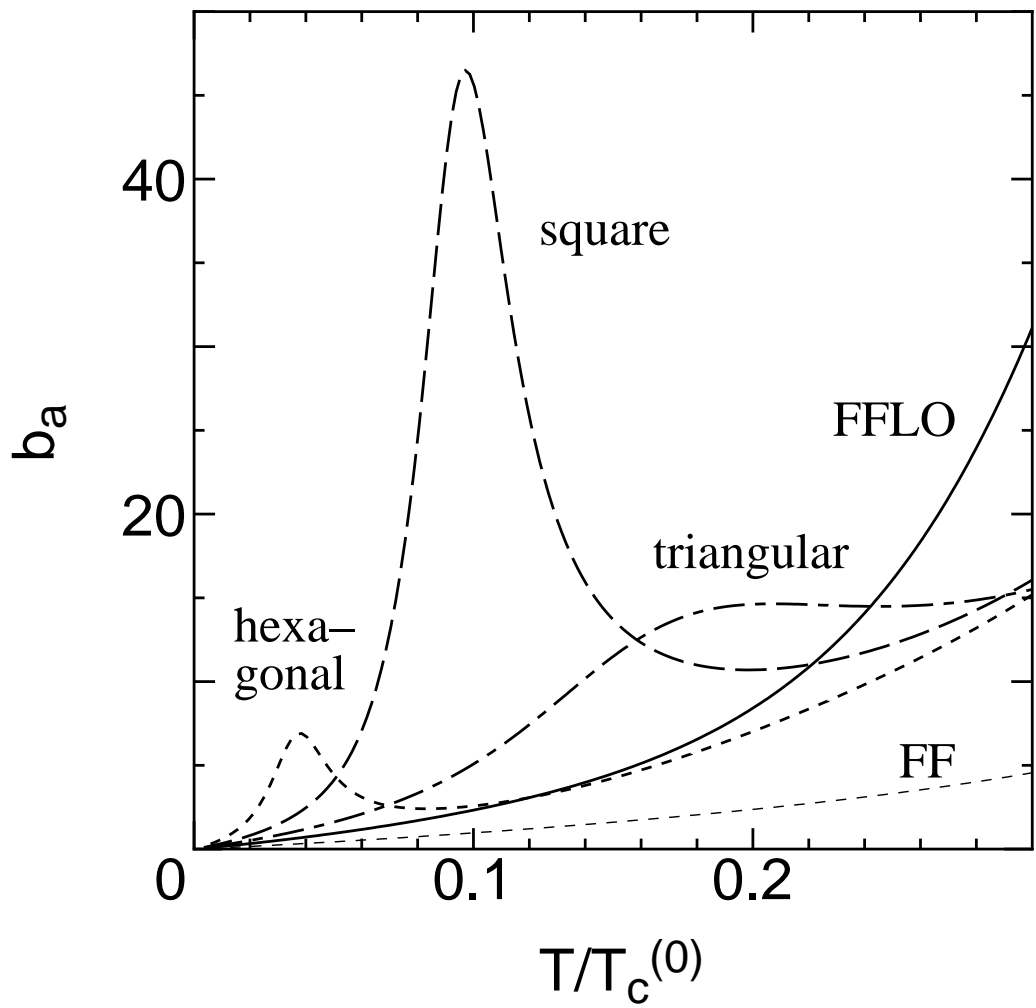
\begin{figure}
\caption{
Factors $b_{\alpha}$ in the free energies of the state (a)-(e) 
for the $s$-wave pairing. 
}
\label{fig:swave}
\end{figure}
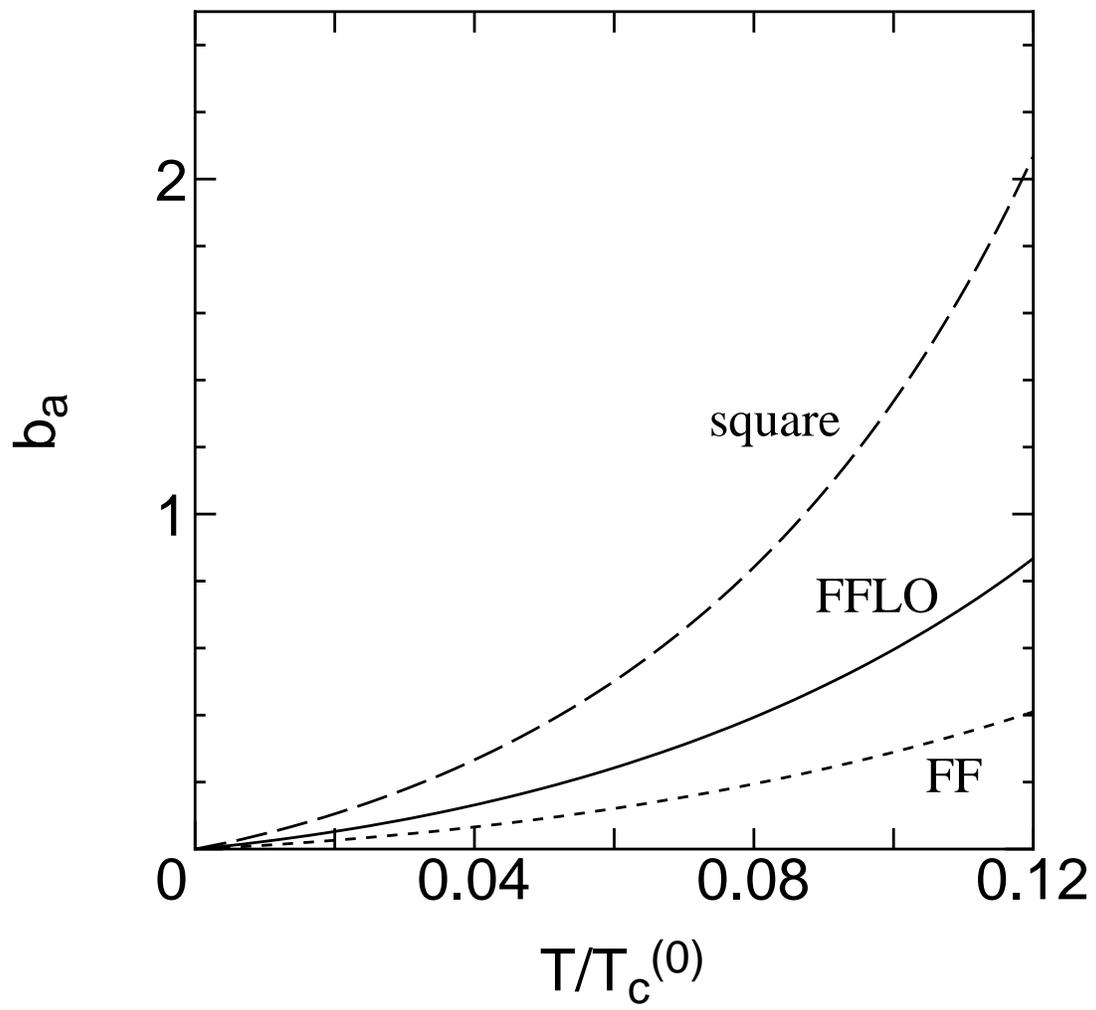
\begin{figure}
\caption{
Factors $b_{\alpha}$ in the free energies of the state (a)-(c) 
for the $d$-wave pairing, when the angle between $\vq$ and $x$-axis 
is equal to $n\pi/2$. 
}
\label{fig:dwavelt}
\end{figure}
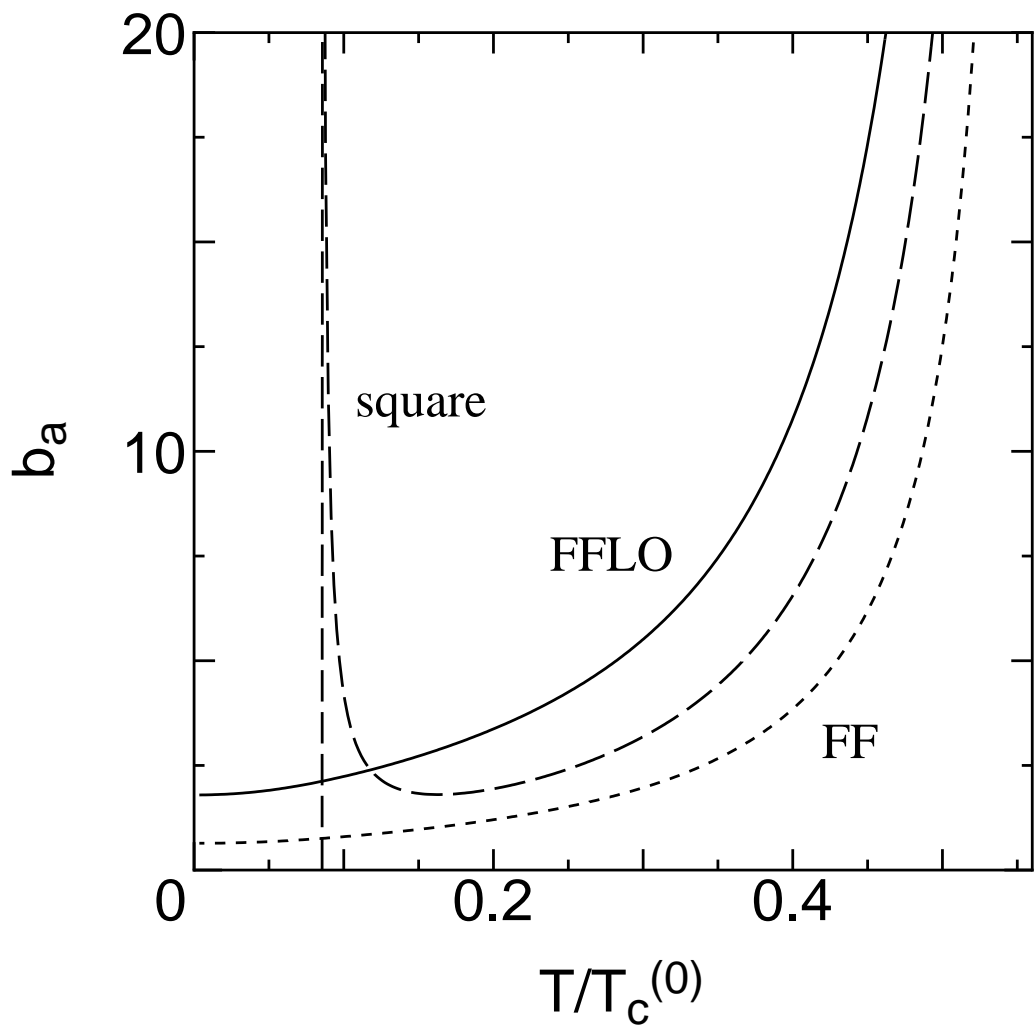
\begin{figure}
\caption{
Factors $b_{\alpha}$ in the free energies of the state (a)-(c) 
for the $d$-wave pairing, when the angle between $\vq$ and $x$-axis 
is equal to $\pi/4+n\pi/2$. 
}
\label{fig:dwaveht}
\end{figure}

\end{document}